# Non-trivial dynamics in a model of glial membrane voltage driven by open potassium pores

Predrag Janjic[1]✉, Dimitar Solev[1], Ljupco Kocarev[1]

[1]Laboratory for Complex Systems and Networks, Research Centre for Computer Science and Information Technologies, Macedonian Academy of Sciences and Arts, Skopje, North Macedonia
✉predrag.a.janjic@gmail.com

**Abstract** - Despite the molecular evidence that the nearly linear steady-state current-voltage relationship in mammalian astrocytes reflects a total current resulting from more than one differently regulated $K^+$ conductances, detailed ODE models of membrane voltage $V_m$ are still lacking. Repeated experimental results of deregulated expressions of major $K^+$ channels in glia, Kir4.1, in models of neurodegenerative disease, as well as their altered rectification when assembling heteromeric Kir4.1/Kir5.1 channels have motivated us to attempt a detailed model incorporating the weaker potassium K2P-TREK1 current, in addition to Kir4.1, and study the stability of the resting state $V_r$. The main question is whether with a deregulated Kir conductivity the nominal resting state $V_r$ remains stable, and the cell retains a trivial, *potassium electrode* behavior with $V_m$ following $E_K$. The minimal 2-dimensional model near $V_r$ showed that certain alterations of Kir4.1 current may result in multistability of $V_m$ if the typically observed $K^+$ currents - Kir, K2P, and non-specific potassium leak are present. More specifically, a decrease or loss of outward Kir4.1 conductance (turning the channels into inwardly rectifying) introduces instability of $V_r$, near $E_K$. That happens through robustly observed fold bifurcation giving birth to a second, much more depolarized stable resting state $V_{dr} > -10~mV$. Realistic time series were used to perturb the membrane model, from recordings of glial $V_m$ during electrographic seizures. Simulations of the perturbed system by constant currents through gap-junctions and transient seizure-like discharges as local field potentials led to depolarization of the astrocyte and switching of $V_m$ between the two stable states, in a downstate – upstate manner. If the prolonged depolarizations near $V_{dr}$ prove experimentally plausible, such catastrophic instability would impact all aspects of the glial function, from metabolic support to membrane transport and practically all neuromodulatory roles assigned to glia.

**Statement of Significance** – The almost linear current-voltage relationship of most glial membranes results from multiple non-linear potassium leaky-pore, or background conductances. The corresponding channel types develop and deregulate independently, some of them asymmetrically – deregulate differently in different $V_m$ ranges. Effect of those deregulations on whole-cell voltage responses has not been treated. We developed a minimal ODE model of voltage dynamics incorporating detailed models of the different potassium currents based on electrophysiological recordings. Parametrically inducing some of the reported deregulations of Kir current in glia resulted in instability of the nominal resting membrane potential and appearence of a second much more depolarized resting state. If prolonged glial depolarizations prove plausible such bistability would change the present beliefs about glial $V_m$ dynamics.

## 1. Introduction

**Membrane voltage control in glia** - The temporal dynamics of membrane voltage ($V_m$) of glial cells, the non-excitable neural cells, is a difficult subject to treat due to their structure-function complexity, and the experimental limitations to precisely measure $V_m$ of their leaky membrane by the standard electrophysiological methods [1]. The classical, more than 50 years old view of the glial membrane as *a passive potassium electrode* [2] reflected the early observations that $K^+$ conductances in astrocytes accounting for the major part of ion conductivity near the resting voltage are of Ohmic nature. It is important to note that those early observations, lacking molecular specificity, have come from studies with a rather narrow experimental focus on the glial role in ion homeostasis, i.e., in the removal of the excess extracellular $K^+$ following sustained neuronal firing. Intensive molecular studies in the last twenty years seriously challenged the view of purely passive electrochemical response capability of glia, upon evidence that their membrane voltage $V_m$ is transiently perturbed by different electrogenic pathways through transporters and ionotropic receptors [3], [4], expressed with certain variations by both neurons and glia. Today we know that glial glutamate and GABA transporters, and even more intriguingly ionotropic glutamate and GABA receptors in some glia cell types [5], do alter glial $V_m$, [6] and therefore introduce forms of feedback control of their respective neuromodulatory functions. Such different control roles of glial $V_m$ translate to a wide range of cellular and systemic functions from glucose transport and neurotransmitter recycling, via responses to mechanical and acidic stress, to the regulation of excitation-inhibition balance in local circuits.

Therefore, with the glial $V_m$ playing both roles: a control variable (*controller*) of different neuromodulatory loops, as well as itself being a neuromodulatory target, the *dynamical stability of glial resting membrane voltage $V_r$* in response to various perturbations remains a central question in quantitative whole-cell modeling of glia. Such a dual effect of $V_m$ as a variable, further qualifies the glial membrane as *a sensor and transducer of $K^+$ changes as a signal*, which perspective has been pushed by pioneers of glial biology [7] but further demands explanatory biophysical models.

The typical, transient perturbations of glial $V_m$ comes either as: (i) transient variations of local field potential which directly polarize the membrane, (ii) varying trans-junctional voltage at the gap-junction connections of heavily interconnected astrocytic cells, or (iii) as transiently altered $[K^+]_o$. Prospective dynamical modeling studies would try to distinguish the specific, causal responses of glial $V_m$ to a specific form of modulation, from the permanent stochastic electrochemical or other perturbations of $V_m$ near the $V_r$.

**Variations of $V_m$ in a steady and perturbed glial membrane -** Experimental studies of astrocytes *in* situ, the most abundant glial cell subtype in the brain, find that their $V_m$ in nominal conditions fluctuates within a narrow range [8], [1]. From several millivolts, in response to nominal fluctuations of the local field, up to $10 \div 20\ mV$ more depolarized from $V_r$ in cases of seizures, other strong synchronous discharges in the neighboring neurons or spreading depolarizations accompanied by high, transiently elevated $[K^+]_o$ [9], [10], or [11] for review. Even though accumulated extracellular $K^+$ depolarizes glia, it is not clear how fast, and how closely glial $V_m$ follows the perturbed $E_K$ when both active and passive transport mechanisms switch in, and what is in turn the sign of the driving force $\Delta V_m = V_m - E_K$ that direct the leaky currents in sustained depolarizations. Typically, the reversal potential of the isolated macroscopic Kir and K2P currents is close, but slightly more positive to the Nernstian reversal potential of potassium, $E_K$, and closer to the cell resting voltage $V_r$ which is typically between $-80\ mV$ to $-70\ mV$ in glial cells, *in situ*. A relevant dynamical model of $V_m$ would thus require incorporating the *minimal*, yet detailed description of the major glial ion conductances active near $V_r$, over several timescales from milliseconds to minutes, under assumption $[K^+]_o$ dynamics is not modeled, i.e., $E_K$ remains a changeable parameter.

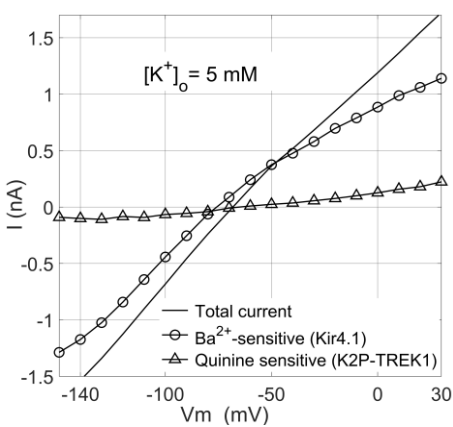

**Figure 1** - **Potassium current profile in isolated astrocytes** – Kir4.1 current isolated as $Ba^{2+}$ sensitive, by application of 0.1mM BaCl (circles / line, N=9) reverses at $V_{rev} \approx 75.35$mV, compared to the total $K^+$ current, black solid line (data from [12]). Of the remaining $K^+$ conductance, K2P current (max. 0.23 nA at +30mV), was isolated as quinine-sensitive, triangles / line. The transcript analysis of the K2P channels suggested a prevalence of K2P2.1 TREK1 isoform [12]. The sum of Kir4.1 and K2P2.1 current accounts for 82% of the total current at -120mV, and 80% at +30mV. Recordings from 9 cells, with a mean effective patch capacitance of 20 pF, have been normalized, averaged, and rescaled in the measured current range.

Figure 1 shows the steady-state current-voltage (I-V) recordings of the total $K^+$ current, as well as the isolated Kir4.1 and K2P-TREK1 currents in our data, obtained under whole-cell voltage-clamp protocol in $N = 9$ freshly isolated astrocytes from mouse hippocampus, bathed in slightly elevated extracellular $[K^+]_o = 5\ mM$ [12]. In steady-state conditions under a physiological $K^+$ concentration gradient, the barium $Ba^{2+}$- sensitive *weakly rectified* Kir4.1 current (circles) dominates the total current profile (solid line). Even though much weaker in resting conditions, the K2P current (triangles) has different kinetics and electrochemical regulation, which requires to have it incorporated into any quantitation of $K^+$ conductivity in astroglia, either experimental or theoretical.

Describing these dominant $K^+$ background currents in isolated astrocytes from mouse hippocampus [12] together with a small $K^+$ leak current is the central modeling effort in this study. The isolated $K^+$ currents operate close to, and therefore define the resting membrane potential $V_r$ in the proposed minimal ordinary differential equation (ODE) model of $V_m$ dynamics. Apart of $K^+$ conductances, no other major conductances have been verified in astrocytes in the voltage range around $V_r$, assuming no osmolar stress which triggers non-negligible $Cl^-$ transients. Co-expressed *Kir* and *K2P* channels account for the major fraction of potassium channels in almost all glial cells in the brain [13], cardiomyocytes and cardiac fibroblasts [14], [15], as well as in various cell types in renal epithelia [16], [17]. With over forty different channel isoforms in total, the co-occurrence of Kir and K2P channels in these, as well as in other tissues [18] is pervasive, resulting in different $K^+$ channel mix and therefore different I-V curves and kinetic properties of the total current.

Parametric analysis of the model reproducing some of the observed biological (de)regulation of those conductances suggests that astrocytic $V_m$ could be prone to multistability in case of deregulation of the major weakly-rectified Kir4.1 current (WR-Kir). Numerical simulations of the model perturbed using realistic recordings from glia *in-situ* [9], illustrated the bistability of the otherwise stable resting state. Although a mix of only $K^+$ conductances may give rise to two membrane resting states, already observed experimentally in cardiomyocytes [14], and described by a simple, generic minimal 2-dimensional conductance-based model [19], it has not been demonstrated so far in the glial membrane or simulated numerically using a glia-specific model.

## 2. Model of astroglial whole-cell $V_m$ dynamics

Detailed electrophysiological studies of glial[1] conductances [12] are rare, and a widely accepted minimal dynamical model of $V_m$ in astrocytes is therefore lacking. Further to the experimental complexity mentioned above, depending on the model system, sometimes additional regulation of these conductances by other environmental stimuli requires specific quantitation.

In what follows we outline the description of experimentally measured steady-state I-V relationships of Kir4.1 and K2P-TREK1 conductances. The nonexistence of a *voltage-sensing domain* - a structure present in other *gated* $K^+$ channels, suggests their rectification properties come from permeability regulation by physiological blocking ions, or by other Coulombic interactions on a pore level [20] resulting in voltage activation. The suggested modeling approaches to either Kir or K2P currents do not follow the Hodgkin-Huxley approach for gated channels but reflect their known biophysics as open pores. The proposed models could be applied *as-are* to a different mix of those channels in different cells or tissues. In the rest of the text, for the cell-specific aspects, without loss of generality, we restrict to an isolated hippocampal astrocyte as our model system [12].

**2.1 Steady-state I-V model of weakly-rectified Kir current -** Since the early quantitative studies, the conductivity of Kir channels under changing $[K^+]_o$, [21] indicate that Kir channels display voltage dependence on the electrochemical driving force $\Delta V_m = V_m - E_K$, rather than solely on the net membrane voltage $V_m$. This implies that proposed models should: (i) allow for variable $[K^+]_o$, and (ii) describe qualitative changes on different timescales, relevant not only for the millisecond $V_m$ dynamics, but for the slower $[K^+]_o$ variations spanning from hundreds of milliseconds to seconds and minutes, as well.

Our central modeling proposition is a detailed description of the I-V relationship of the *weakly-rectifying* potassium Kir current (WR-Kir), rather than straightforwardly assigning the Hagiwara model of strongly inwardly-rectifying channels in egg cells [21]. The Hagiwara model is usually chosen under the assumption that $V_m$ is very tightly regulated round the negative resting potential of $V_r \approx -80 mV$, which is oversimplification because astrocytic $V_m$ is exposed to perturbations that significantly depolarize the cell into a $V_m$ range where the curves of inwardly and weakly rectified Kir current markedly differ. To fill this gap requires describing quantitatively the outward conductivity of WR-Kir channels, which represent a comparable fraction to the inward component, Fig. 1 (circles).

---

[1] In the rest of the text, we will interchangeably use the terms *glia, astroglia* or *astrocytes* for the population of the astrocytic cells. Where reference is made to other glial cell types, those are named specifically.

The key assumption is that the steady-state WR-Kir current, Fig. 1 (circles), could be *dissected into two additive components*: a) inwardly rectified Kir component $I_{Kir}$, and b) outwardly rectified, or *residual Kir component* $I_{res}$, so that for the total isolated $Ba^{2+}$-sensitive Kir4.1 current we can write:

$$I_{Kir} = I_{inw} + I_{res} \qquad (nA). \qquad (1)$$

The two unidirectional fluxes result from a different manifestation of a *voltage-dependent block* of the pore by polyvalent cation which in turn result in different outward and inward permeation. Structure-function studies of Kir channels have shown a large extent of molecular similarity between the strongly and weakly rectifying channels, where a single point mutation dominantly impacts unidirectional permeation in open pores [22], [23]. We describe the permeation on a *short-pore model* of $K^+$ pore, formulated using a simplified one-dimensional reaction coordinate. It is generic for all $K^+$ open pores, where the cytoplasmic domain of the channel is ignored as not critical for describing the permeation, Fig. A1. See *Appendix 1* for a detailed description of the short-pore and further biophysical reasoning on the legitimacy of modeling two separate, additive fluxes in the weakly rectifying Kir pore, see *Appendix-1*. In the rest of the text charge will be represented in molar units of $F, (C/mol)$ and the energy in the energy profiles will be normalized by $RT = 0.593\ kcal/mol$ at $T = 298\ K$ accordingly and expressed in *RT units*.

**Inwardly rectified WR-Kir flux** results from a negative driving force, $\Delta V_m < 0$, where the blocking ions, $Mg^{2+}$ or other polyvalent cations are kept within the water cavity (Fig.A1.B) of the short-pore by Coulombic forces of the negative residues of transmembrane helices (Fig. A1.A). Such ion-crowding of permeant and blocking ions (Fig. A1.B) results in an inward pseudo gating of the WR-Kir pore producing an equilibrium probability of open, inwardly conducting pore described by Boltzmann term, as originally introduced by the Hagiwara model [21]:

$$I_{inw} = g_{Kir-inw}(V_m - E_{Kir}) \qquad (nA) \qquad (2)$$

$$g_{Kir-inw}(\Delta V_m) = \frac{A\ \bar{g}_{s-inw}\sqrt{[K]_o}}{1 + \exp(-z_{inw}(\Delta V_m - \Delta V_{12})/v_s)} \qquad (\mu S) \qquad (3)$$

$$\Delta V_m = (V_m - E_{Kir}),\ \Delta V_{12\_inw} = (V_{12-inw} - E_{Kir}) \qquad (mV),$$

where $\bar{g}_{s-inw}$ represents the maximal value of the $g_{Kir-inw}$ obtained as a slope of the linear segment in the recorded I/V curve negative of $E_K$, $V_{12\_inw}$ is the voltage at half-maximal conductance, $z_{inw}$ represents the effective charge valence of permeant ions in the short-pore, $E_{Kir} = -76\ mV$ is the reversal voltage of the isolated Kir4.1 current, and $A = 1.0\ mM^{-1/2}$ keeps the conductance of $\mu S$. Figure 2A, the solid black curve, shows a fitted sigmoid function, Eq. 3, to numerically differentiated Kir4.1 I-V data, from Fig. 1 (blue curve, for $[K^+]_o$ = 5mM), for $V_m$ more negative than $E_{Kir}$. Such numerically obtained series physically represents a slope conductance, whereas the values more negative to -110mV (blue crosses) are discarded due to slope deviation (Fig.1) originating from a known effect of voltage-dependent pore block by external $Na^+$ ions [24], [25]. The fitting gave $A \cdot \bar{g}_{s-inw}\sqrt{[K]_o} = 20.6\ nS$, $V_{12\_inw} = -53.5\ mV$ and $z_{inw} = 1.638$, producing the blue curve in Fig. 2B. The effective charge $z_{inw}$, was in accordance with published charge estimates for the voltage dependence of block by small ions, for a summary see [26]. The slope factor $v_s \equiv RT/F = 25.7\ mV$, corresponding to room temperature of $T = 298.13\ K$ is kept constant throughout the text.

**Outwardly rectified Kir4.1 flux** results from outward pseudo gating for a more positive driving force $V_m > -60mV$ (Fig. A1.C), where outwardly directed driving force pushes partly or fully dehydrated blocking ion towards the entry of the selectivity filter (SF), see *Appendix 1.* Such electrostatic, flickering block modulates ion association to the pore and the outward permeation.

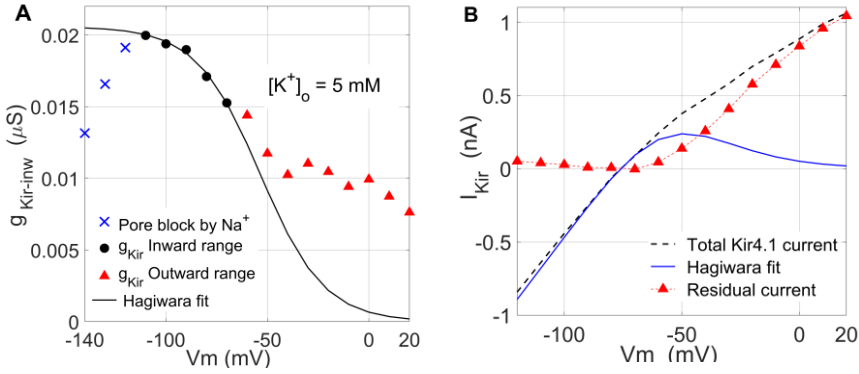

**Figure 2**- **Separation of WR-Kir current in two unidirectional fluxes** – **(A)** Fitting the conductance $g_{Kir-inw}$ by the Hagiwara model for $V_m \leq -60 mV$, black line, with the crosses excluded. **(B)** The *residual outward* component, eq. (4), red triangles, obtained by subtracting the modeled inward current (blue), Eq. 2, from the total recorded $Ba^{2+}$ sensitive Kir component (dashed line). Blue crosses in (A), and the small non-zero values carried over by the subtraction in (B) (red) for $V_m < -100\ mV$, result from pore block by $Na^+$.

To obtain numerical series for the residual, outward Kir4.1 current $I_{res}$ (4) from the recordings, we subtracted the portion *described by the* Hagiwara model in the whole $V_m$ range, Eq. 2 and Eq. 3, fitted near $V_r$, Fig.2B, blue line, from the total Kir current, red triangles in Fig.2B.

$$I_{res} = I_{Kir} - I_{inw} \quad (nA). \tag{4}$$

**Non-constant field Goldman-Hodgkin-Katz (GHK) model of WR-Kir outward current** - Without a voltage gate as a structural determinant of gating in Kir and K2P pores, other features like physiological block by cytoplasmic cations may produce pseudo gating by charge movement within the membrane or alter the Coulombic and/or structural forces defining open-pore permeability *within the single global state* of the channel.

We arrive at the proposed model of the residual current using the following assumptions:

- *GHK formalism without constant-field assumption* describes the outward conductivity, using Eq. 5 as a general form of GHK current equation before introducing the constant-field assumption [27], chapter 13.13. Instead, a simple, *quadratic potential energy profile* of an elastic force $U(x)$ introduces a parabolic energy barrier in the energy profile of an open Kir pore, Fig.3A, with the voltage drop over the pore added as linear, $V_m(x) = x\Delta V_m$. The $U_{max}$ potential represents the peak of the *association* barrier to the permeant $K^+$ ion at the entry of hydrophobic SF in $RT$ units. The barrier is centered at transition state $x_T$, while the positive $w$ is the *force constant* of the elastic restoration force of Coulombic nature. For the charge in molar equivalents and the concentrations in $[mM]$, we will describe the outward current using:

$$I_{res} = zFD \frac{[K^+]_i - [K^+]_o e^{-\frac{zFV_m}{RT}}}{\int_a^b e^{U(x)-zFV_m(x)/RT}\, dx} \quad (nA), \tag{5}$$

$$U(x) = U_{max} - w(x - x_T)^2 \quad (RT\ units).$$

We can extend the interval $(a, b)$ to $(-\infty, +\infty)$ because the energy profile falls sharply outside the short pore. Depending on the description of $U_{max}(V_m)$ the location of $x_T$ could change, resulting in an asymmetrical shape of the barrier, sketched in Fig. 3A, dashed line, compared to the symmetrical case in Fig. 3B (barrier not shown). Different positions of the transition state, $x_T = l_{sp}/2$ in Fig. 3A, and $x_T = l_{sp}/3$ in Fig. 3B simply illustrate different model assumptions for the barrier position.

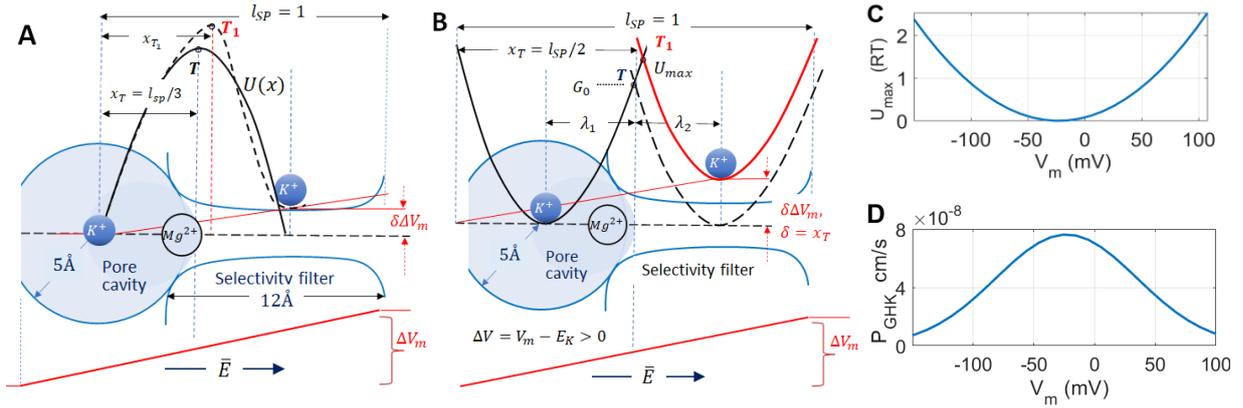

**Figure 3 – Reaction coordinate in the model of WR-Kir outward permeation** – **(A)** Quadratic energy barrier profile centered at the transition state $x_T$ used in Goldman-Godgkin-Katz description of the outward flux of Kir4.1 current. Here the reaction coordinate starts at the center of the cavity, and the $\delta V_m$ fraction of the voltage drops over the reaction coordinate, with a full length $l_{SP} = 1$. **(B)** Sketch of the Marcus free energy formalism defining voltage-dependence of $x_T$ and the barrier height $U_{max}$. Voltage-dependent shift of $x_T(V_m)$ from $T$ to $T_1$ and the selection of $\lambda_1$ and $\lambda_2$ positions of the energy wells in (B) produce either a symmetric (solid line) or skewed barrier at the top (dashed line) in (A). Here, compared to (A), the reaction coordinate spreads over the whole short-pore, and $x_T = 1/2$ places the peak of the symmetrical barrier at the entry of SF. **(C)** and **(D)** peak behavior of $U_{max}$, Eq. 7, and non-constant GHK permeability (9) at $V_C = -23\ mV$ which will produce maximal outward current $I_{res}$. The peak of $P_{GHK}$ in (D) suggests the facilitating Coulombic effect of a polyvalent blocker on the outward permeation is maximal at $V_C$ above which it gradually turns into full pore occlusion.

- **Partial voltage-dependent pore occlusion by cytoplasmic cations activates the pore in the outward direction –** In this model voltage-driven localization of polyvalent cytoplasmic cations in the cavity produces a partial, flickering pore occlusion, Fig. 3A, which increases the Coulombic repulsion on the permeant ion associated with the SF. Such partial blocking serves as a plausible modulatory mechanism, with the pore remaining in a single open conformation. Therefore, both, the pseudo-gating on pore-level and the macroscopic current rectification are attributed to the voltage dependence of the block. This defines a voltage-dependent probability of an activated pore with an *effective, equivalent charge* $z_B > 1$, reflecting the valence of the blocking ion ($Mg^{2+}$ or positively charged polyamines), not necessarily implicating a multi-ion pore. We illustrate this blocking configuration along the reaction coordinate of the short-pore model on the sketch in Figure A1.C, as a displacement of the blocking $Mg^{2+}$ ion from S6 (center of the cavity) towards the S5 ion coordination site (SF entrance). To formulate the modulatory effect of blocking as a probability of *outward activation* in a non-gated channel, we use the Boltzmann equation in a classical way [27], chapter 1.12:

$$p_{out} = 1/1 + e^{-z_B(V_m - V_{12-out})/v_s}. \qquad (6)$$

Since both, the Hagiwara model of Kir, as well as more recent permeation studies on K2P channels [28], suggested electro-chemical gating in $K^+$ pores, with effective driving force $\Delta V_m = V_m - E_K$, applying it within Boltzmann equation cancels out the reversal potential $E_K$ in Eq. 6, as it also appears in the half-activation voltage $\Delta V_{12-out} = V_{12-out} - E_K$ (same as in Eq. 3). The slope factor $v_s$ is the same defined within the model for inward flux.

- **Quadratic dependence of barrier peak $U_{max}$ on voltage saturates the outward flux** - We are adding nonlinear dependence of Marcus form, Eq. 7, to introduce a simple form of Coulombic contribution of the voltage-dependent block in $U_{max}$ according to [27], see chapters 7.6 and 7.8:

$$U_{max}(V_m) = G_0(\lambda - x_T)^2 = G_0\left(\lambda - \frac{z_B\delta(V_m - E_K)}{4\lambda v_s G_0}\right)^2; \quad \lambda_1 = \lambda_2 \equiv \lambda, \quad (RT\ units), \qquad (7)$$

which can describe the *saturation of outward flux* for very positive applied voltages, Fig.4. Figure 3B shows qualitatively the geometry of the Marcus energy landscape and $U_{max}(V_m)$ over a simplified, one-dimensional reaction coordinate. The peak of the barrier $U_{max}$ at the intersection of the two parabolic wells, $2\lambda_1$ and $2\lambda_2$ wide, defines the transition state $x_T$. It separates the left energy well where the permeant ion dwells before the outwardly directed field drives it to the SF and the right well where the ion is associated to the pore. The critical parameter influencing $U_{max}$ voltage dependence is the fraction of the linear voltage drop $\delta = 1/2$ that falls under the energy profile, reducing it to $\delta \Delta V_m{}^2$. Figure 3C illustrates the quadratic profile of $U_{max}(V_m)$ with the minimum attained at a critical voltage $V_C = -23 \, mV$. The nonlinear $U_{max}(V_m)$ dependence is significant for capturing the outward WR-Kir current saturation with voltage [29], [30], [20] which is not possible with a constant or linearly dependent $U_{max}$. The prefactor $G_0$ in RT units represents the height of the constant barrier due to hydrophobic repulsion in the absence of electrochemical driving force.

Equation 7 is based on the Marcus chemical kinetics theory describing the charge transfer in chemical bonding, using reaction rate models [31], which has meanwhile gained a wider acceptance in treating quasi-equilibrium changes in proteins, for review see [32]. We stress that we use here qualitative features of very simplified one-dimensional reaction coordinate in the Marcus theory, *to describe* macroscopic properties, like the saturation in the voltage dependence of block in measurements of macroscopic whole-cell currents, rather than to construct a model for predicting permeation rates or estimating the microscopic parameters of the pore, like $z_B$ or $G_0$ [33].

Under the above assumptions, by solving the Gaussian integral in the nominator of Eq. 5 and simplifying the solution by setting a fixed position $x_T = 1/2$ [27], for the outward residual component $I_{res}$ of isolated Kir4.1 current we obtain:

$$I_{res} = p_{out} \, zFP_{GHK} \left( [K^+]_i e^{\frac{zV_m}{2v_s}} - [K^+]_o e^{-\frac{zV_m}{2v_s}} \right) \quad (nA). \quad (8)$$

$x_T = 1/2$ places the transition state and $U_{max}$ at the entry of the SF where the blocking ion should be localized for positive $\Delta V_m$, which is close to 1/2 of the length of the short pore with the cavity approximately 10Å in diameter, and the SF length typically approximated to $12 \div 15$Å, depending on the structure-function assumptions. The simplification removes $x_T(\Delta V_m)$ shift which if present, for illustration, for $\Delta V_m = v_s$, $\lambda_1 = \lambda_2 = 1/4$, with $z_B = 1.6$ and $G_o = 6.6 \, RT$ obtained by fitting Eq. 8 to the series obtained by Eq. 4, shifts $x_T$ from 0.5 to 0.56.

Having departed from the classical Nernst-Planck assumptions in solving Eq. 5 for a macroscopic description of $I_{res}$, by using Eq. 7 to add modulation of permeability by voltage-dependent block, $P_{GHK}$ introduced in Eq. 8 absorbs the resulting prefactors and implicitly defines voltage-dependent permeability [27]:

$$P_{GHK}(V_m) = \sqrt{\frac{w}{\pi RT}} D \, e^{-\frac{U_{max}(V_m)}{RT}} = P_K e^{-\frac{U_{max}(V_m)}{RT}} \quad (cm/s). \quad (9)$$

Here $w$ is the same force factor from the harmonic potential $U(x)$ in (5), and $D$ is the classical, constant diffusion coefficient, both absorbed within the constant, voltage-independent $K^+$ permeability $P_K$ multiplied by a factor of Arrhenius activation form with voltage-dependent barrier $U_{max}(V_m)$ introduced in Eq. 8. The zero RT minimum of $U_{max}(V_C)$, Fig. 3C, should be interpreted as a zero of the *voltage-dependent potential energy profile* at which we observe the maximal permeability $P_{GHK}$ reducing to the constant value of $P_K = 7.63 \times 10^{-8}$ cm/s in Nernst-Planck electro-diffusion, Eq. 9, Fig. 3D. The estimate is within the same order of magnitude as the ranges in the early studies of rectification in potassium channels, see [34] for a review. For references where explicit nonlinear dependence of GHK permeability has been used, mostly in models of neuronal calcium channels, see [35].

---

[2] $\delta$ represents a dimensionless fraction of the length of the pore $0 \geq \delta \geq l_{SP} \equiv 1$, or the "electrical" length of the pore.

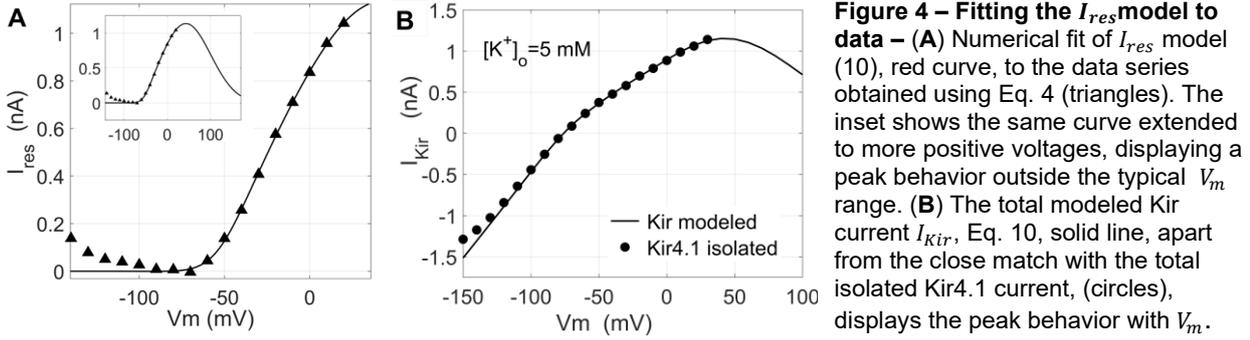

**Figure 4 – Fitting the $I_{res}$ model to data –** (**A**) Numerical fit of $I_{res}$ model (10), red curve, to the data series obtained using Eq. 4 (triangles). The inset shows the same curve extended to more positive voltages, displaying a peak behavior outside the typical $V_m$ range. (**B**) The total modeled Kir current $I_{Kir}$, Eq. 10, solid line, apart from the close match with the total isolated Kir4.1 current, (circles), displays the peak behavior with $V_m$.

For the final form of the Kir current model to fit the isolated Kir4.1 current recordings, we get:

$$I_{Kir} = I_{inw} + I_{res} \quad (nA)$$

$$I_{inw} = g_{Kir}(V_m - E_K) = \frac{A\,\bar{g}_{s-inw}\sqrt{[K^+]_o}}{1 + e^{-z_{inw}(V_m - V_{12})/v_s}}(V_m - E_K) \quad (nA) \quad (10)$$

$$I_{res} = p_{out}zFP_K e^{-U_{max}/RT}\left([K^+]_i e^{\frac{zV_m}{2v_s}} - [K^+]_o e^{-\frac{zV_m}{2v_s}}\right) \quad (nA).$$

Fitting $I_{res}$ model (10) with $U_{max}$ given by Eq. 7 to the residual outward current data obtained by Eq. 4, using nonlinear least-squares error, gives the I-V graph in Figure 4A. The peak behavior in Eq. 7 produces a peak behavior of $I_{res}$ in a more positive, non-physiological $V_m$ range for glia, Fig.4B. Such shape has already been reported in a study of a whole rodent hippocampus [29], with the curve shifted to more negative voltages.

All the parameter values, the fixed ones, and those obtained by fitting are listed in Table 1. The table values of $I_{res}$ fitting parameters: $P_K, V_{12-out}, G_0, z_B,$ and $z$, to be later used in the parameter analysis of the model, were obtained as averages of ten curve fitting attempts with one of them varied within $\pm 10\%$ of the initial value in each run while the other four were recalculated by the NLSQ curve fitting. The blocking ion charge valence was bounded within $1.5 \leq z_B \leq 2.0$. The goodness of fit $R^2 \geq 0.98$ was achieved in all attempts.

**2.2  Steady-state I-V model of K2P-TREK1 current –** Studies isolating the weaker astrocytic currents after isolating Kir currents are rare, due to experimental issues with leaky astrocytic membrane, and therefore K2P currents are rarely quantified [12], [36]. Figure 5A shows pharmacologically isolated K2P current, using *quinine*, at $[K^+]_o = 5\ mM$ (black circles), as well as at drastically elevated external potassium at $[K^+]_o = 50\ mM$ (blue dots) in our recordings from [12]. Both I-V recordings were fitted (solid lines) with the standard GHK current equation [37], even though the permeation studies suggest multi-ion occupancy of the pore and some form of pseudo-gating [28]. Adding a small $Na^+$ permeability [37], $P_{Na}/P_K = 0.06$, within nominally non-conductive range (below $-50\ mV$) improves the fit at $5\ mM$, Fig. 5A, dotted line. That effect of $Na^+$ on passive $K^+$ conductivity in glia is known [38], but we didn't use the correction within the numerical simulations due to a lack of recordings to verify $P_{Na}/P_K$ ratio for notably elevated $[K^+]_o$ above 5 mM.

The K2P currents in this study [12] originate primarily from the K2P-TREK1 sub-population of K2P channels. TREK1 channels are polymodal transducers of different cellular stimuli [39], [40], like mechanical pressure or pH shifts that transduce into changes of $V_m$, so that any distinction between steady-state and activated state of these and most of others K2P channels requires careful consideration, both experimental and biophysical. Working with steady-state recordings from astrocytes isolated in a bath assures that obtained I-V curves represent cells in steady conditions, non-transducing an osmolar stress or pH variations, which is implicitly assumed in a baseline model.

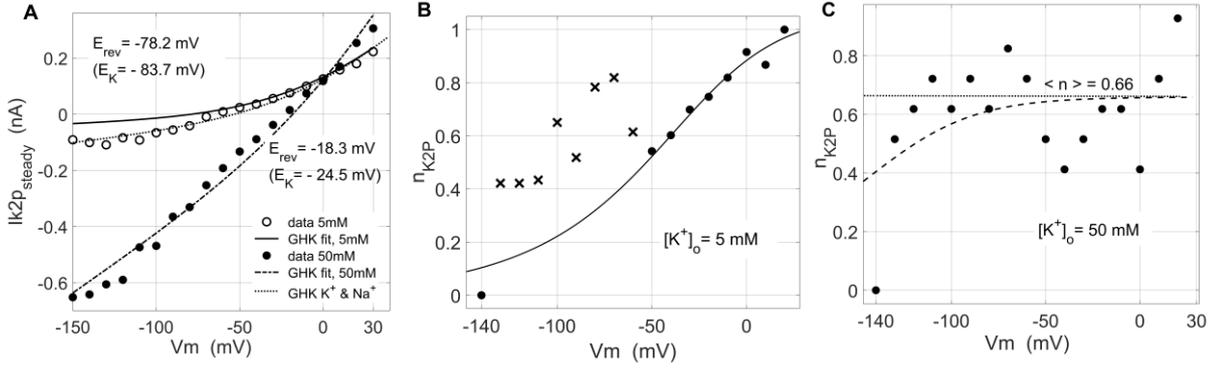

**Figure 5 - Steady-state I-V relationships and activation kinetics of K2P-TREK1 currents in astrocytes – (A)** Steady-state I-V characteristics of K2P-TREK1 currents fitted with the Goldman-Hodgkin-Katz current equation, for $[K^+]_o$ = 5mM (circles/solid line) and 50mM (filled circles/dashed line). The dotted curve shows the fit corrected with a small $Na^+$ permeability being added ($P_{Na+}/P_{K+} = 0.06$, see text). **(B)** and **(C)** Approximation of the activation kinetics $n_{K2P}(V)$ of K2P current for $[K^+]_o$ = 5mM and 50mM respectively, obtained as $(g_{slope} / \bar{g}_{max})^{\wedge}(1/k)$, $k = 2$, see text. At 5mM, in (B), the $n_{K2P}$ activation estimates for $V_m$ positive to $-50 mV$ (filled circles) were fitted with Boltzmann first-order activation kinetics, ignoring more negative $V_m$ values where very small currents produced erratic estimates of $g_{slope}$, obtained by numerical differentiation (crosses). **(C)** The loss of voltage-dependent activation $n_{K2P}$ (as seen in B) at $[K^+]_o = 50\ mM$ was modeled as a gradual $[K^+]_o$- *dependent shift* of the $n_{K2P}$ curve (see Eq. 12b) to the left practically removing the sigmoidal voltage dependence and reducing $n_{K2P}$ to fluctuations around a mean, constant activation $\langle n_{K2P} \rangle = 0.66$, for voltages in the activated range for K2P current.

Apart from the outward voltage activation (verified also in a symmetrical $K^+$ concentrations), K2P channels show also some form of *flux-dependent activation* component coming from multi-ion occupancy of the pore by the permeant $K^+$ ions [28]. The following two dynamical properties in voltage-dependent activation of K2P-TREK1 current are essential for arriving at the full dynamic model of $V_m$:

- **Activation kinetics** - At both concentrations, the whole-cell currents showed activation kinetics with $\tau_{K2P} \approx 3\ ms$, constant over the whole $V_m$ range. While at $5\ mM$, above $-50\ mV$ we could fit a Boltzmann sigmoidal, voltage-dependent activation, Fig. 5B black dots, at drastically elevated, $[K^+]_o = 50 mM$ there is no visible voltage dependence, Fig. 5C, described with a constant average activation of $\langle n_{K2P} \rangle = 0.66$.
- **Gating mode change** – At elevated external $K^+$, Fig. 5A, the dominating *electrochemical activation* [28] is changing the I-V curve, resembling more that of the leaky K2P-TWIK channels with almost linear dependence, satisfying the GHK current equation and reversing close to the Nernst potential.

Along with these observations we model the astrocytic K2P-TREK1 currents using the GHK current model [37], extended with a complex electrochemical activation $n_{K2P}$ – coupling the voltage-dependent activation and electrochemical driving force into a mechanism named *ion-flux gating* [28]:

$$I_{K2P} = n_{K2P}^k I_{K2p-GHK}$$
$$= n_{K2P}^k P_{K2P} \frac{F^2 z_{K2P}^2}{RT} V_m \frac{([K^+]_i - [K^+]_o \exp(-z_{K2P} V_m/v_s))}{(1 - \exp(-z_{K2P} V_m/v_s))}\ (nA). \tag{11}$$

Allowing for variations of $K^+$ concentrations we cannot keep the intrinsic permeability $P_{K2P}$ constant [41], contrary to the classical assumptions based on the solubility-diffusion theory [26], [42]. To add a simple $[K^+]_o$ dependence in $P_{K2P}$ we interpolated the fitted values for $P_{K2P}$ at 5mM and 50mM with an increasing and slowly saturating $K^+$ dependence, Eq. 12a. The GHK fit at 5mM suggested $P_{K2P}^b = 1.24 \times 10^{-8} cm/s$ for the baseline permeability, within the order of magnitude with those reported in the early studies of electro-diffusion in $K^+$ channels, [43]. As an additional sanity check, the ratio $P_K/P_{K2P} = 6.15$ of the constant permeability values obtained by fitting the GHK models of the Kir $I_{res}$ and $I_{K2P}$, is reasonably close to 5.3, the ratio of their maximal measured values at 30mV, Fig. 1.

Fitting of the sigmoid in Fig.5B to $\left(g_{slope}/\bar{g}_{max}\right)^{(1/k)}$ graph obtained by differentiation for both concentrations was done with $k = 2$, following as a guiding value the average ion occupancy of the SF with 2.2 $e_0$ elementary charge units across different K2P channel subtypes, estimated in [28]. Small voltage offset ($V_m - V_{ofs}$) not exceeding 5 $mV$ was needed to stabilize numerically the nonlinear LSQ fitting of (11). The same value $z_{K2P} = 1$ was used at both $K^+$concentrations.

The nominal, baseline external concentration $[K^+]_o^b$ was kept at 2.5 $mM$ in Eq. 12a. To correct the $n_{K2P}$ activation for variable concentrations two interventions are needed: (i) decreasing trend of the maximal activation with $[K^+]_o$, Eq. 12b, where the $[K^+]_o/[K^+]_i$ term successfully captures the trend, and (ii) shifting the $V_{12-K2P}$ (12c) and the whole $n_{K2P}$ sigmoid to the left (into non-physiological hyperpolarized range) so that the plateau of $\langle n_{K2P}\rangle = 0.66$ covers the most of physiological $V_m$ range - practically removing voltage-dependent activation at high $[K^+]_o$, Fig. 5C. More recordings on different $[K^+]_o$ are needed to verify whether the suggested corrections in Eq. 12b and Eq. 12c have more general biophysical utility. The relative shift of Nernstian nature in (12c) required a scaling factor $S = 1.7$ to get the plateau-like in Fig. 5C. Similar proportionality of $V_{12}$ shift was observed in activation curves of the human K2P-TREK1 channels expressed in *Xenopus* oocytes [28].

$$P_{K2P} = P_{K2P}^b(1 + 0.85 \log_{10}([K^+]_o/[K^+]_o^b)) \quad (cm/s), \quad (12a)$$

$$n_{K2P}(V,[K^+]_o) = \frac{1 - [K^+]_o/[K^+]_i}{(1 + \exp(-z_{K2P}F(V - V_{12\_K2P})/RT))}, \quad (12b)$$

$$V_{12-K2P}([K^+]_o) = V_{12-K2p}^0 - S\, v_s \ln([K^+]_o/[K^+]_o^b) \quad (mV). \quad (12c)$$

Equation 11 fuses (i) the modified voltage-dependent and $[K^+]_o$- corrected $n_{K2P}$ activation, Eq. 12b, describing an arbitrary K2P channel population, and (ii) the nonlinear Goldman-Hodgkin-Katz electro-diffusion describing the macroscopic steady-state current of the whole, voltage-clamped cell. The GHK current model extended with activation kinetics has already been used in descriptions of neuronal voltage-gated calcium channels [44].

**2.3    Full model of glial $V_m$ dynamics** – The dynamics of the astrocytic membrane voltage $V_m$ is described by the differential equations Eq. 13, where to Kirchhoff's current law a single kinetics equation has been coupled, that of K2P channel activation $n$.

$$\begin{aligned} C_m\, dV_m/dt &= -(I_{Kir} + I_{K2P} + I_{leak}) + I_{ext} \quad (nA) \\ dn/dt &= (n_{K2P}(V_m) - n)/\tau_{K2P}. \end{aligned} \quad (13)$$

An unspecific, potassium Ohmic leak current is added as $I_{leak} = g_{leak}(V_m - E_K)$ with the conductance not exceeding 15% of the chord conductance of total glial $K^+$current in the I-V plot, proportional to the current that remains after isolating Kir and K2P currents, Fig 1.

The external inputs, resembling external currents to an astrocyte *in situ*, are represented as:

$$\begin{aligned} I_{ext} &= I_{gjc} + I_{lfp} \quad (nA) \\ I_{gjc} &\sim \sum_i g_c^i(V_m^i - V_m) = const. \quad (nA) \quad i = 1,\ldots N_{neighbors} \\ I_{lfp}(t) &= g_{lfp}V_{lfp}(t) \quad (nA), \end{aligned} \quad (14)$$

where $I_{gjc}$ brings in the contribution from all neighboring astrocytes connected via Ohmic gap junction connections (GJCs) and will be varied as a parameter, while $I_{lfp}(t)$ represent the transient, variable depolarizations of the membrane coming from the changing local field potential immediate to the astrocytic membrane, resulting from prolonged neuronal spiking episodes.

| Parameter | Units | Description | Value, default | Comment |
|---|---|---|---|---|
| $T$ | $K$ | Temperature | 298 | Room temperature, 25°C in [12] |
| $F$ | $C/mol$ | Faraday constant | 96485 | |
| $R$ | $J/mol \cdot K$ | Universal gas constant | 8.314 | |
| $v_k$ | $mV$ | Slope factor, $RT/F$ | 25.7 | at $T = 298°K$, constant in all models and simulations |
| $[K^+]_o^{nom}$ | $mM$ | Baseline $[K^+]_o$ in extracellular space | 2.5 – 5 | 5mM used in I-V fitting, varied within param. analysis |
| $[K^+]_i$ | $mM$ | Astrocytic $[K^+]_i$ | 130 | |
| $C_m$ | $pF$ | Astrocytic membrane capacitance | 20 | Mean effective capacitance of the patch, whole-cell |
| $\bar{g}_{s-inw}$ | $\mu S$ | Max. Kir inward slope conductance | 0.00917 | Hagiwara model of Kir current, Eq. 3 |
| $A$ | $1/mM^{1/2}$ | Dim. constant, correcting for $\sqrt{[K^+]_o}$ | 1 | Dimensionality correction |
| $V_{12-nom}$ | $mV$ | Half-activation voltage $V_{12}$ at $[K^+]_o^{nom}$ | -53.5 | Hagiwara model of Kir current, Eq. 3 |
| $z_{inw}$ | | Effective charge valence, $I_{Kir}$ | 1.638 | Hagiwara model of Kir current, Eq. 3 |
| $z$ | | Unitary valence of permeant $K^+$ ions | 1.0 | |
| $E_{rev-Kir}$ | $mV$ | Rev. voltage of isolated Kir4.1 current | -76 | Average exp. measured reverse voltage |
| $P_K$ | $cm/s$ | Constant permeability of $K^+$ in pores | 7.63e-08 | $K^+$ permeability in classical Nernst-Planck formalism |
| $V_{12-out0}$ | $mV$ | Half-activation voltage of $I_{res}$ | -51.4 | Boltzmann term in $p_0$, Eq. 6 |
| $z_B$ | | Effective charge valence of the blocker | 1.6 | $Mg^{2+}$ or polyamine blocking ion |
| $G_0$ | RT units | Voltage-independent entry barrier | 6.6 | Barrier height at $x_T$, Marcus term, Eq. 7 |
| $l_{SP}$ | | Normalized, unitary length of the pore | 1 | Short pore model, Fig. A1 |
| $x_T$ | | Fractional distance of transition state | 1/2 | Position relative to the beginning of the pore |
| $\lambda$ | | Half-width of the energy wells | 1/4 | Marcus term, Eq. 7 and Fig. 3 |
| $\delta$ | | Fractional, electrical length of the pore | 1/2 | Fraction of $l_{SP}$ over which $V_m$ drops, Marcus term, Eq. 7 |
| $V_{12-K2p}^0$ | $mV$ | Half activation voltage $V_{12}$ at $[K^+]_o^{nom}$ | -20.5 | $I_{K2P}$ model, $n_{K2P}$ activation, Eq. 12c |
| $k$ | | k-th power in K2P activation kinetics | 2 | Hodgkin-Huxley formalism, Eq. 11 |
| $S$ | | Scaling parameter in $n_{K2P}$ activation | 1.7 | Adjusts the shift of $n_{K2P}$ with $[K^+]_o$, Eq. 12c |
| $z_{K2P}$ | | Charge valence for K2P pores, fixed | 1.0 | $I_{K2P}$ model, Eq. 11 and $n_{K2P}$ activation, Eq. 12b |
| $\tau_{K2P}$ | $ms$ | Activation time of K2P-TREK pore | 3.0 | From exp. data [12], Eq. 12b |
| $P_{K2P}^b$ | $cm/s$ | K2P channel permeability, basal | 1.24e-08 | $I_{K2P}$ model, GHK description, Eq. 12a, |
| $g_{leak}$ | $\mu S$ | Glial leak conductance, Ohmic | 0.001-0.002 | Non-specific leak, glial voltage dynamics, Eq. 13 |
| $I_{ext}$ | $nA$ | External current to the astrocyte | 0.2 - 0.4 | Glial voltage dynamics, Eq. 13 |
| $g_{lfp}$ | $\mu S$ | Transfer conductance in resp. to $V_{lfp}$ | 0.01-0.016 | Adjusted to the $I_{ext}$ ranges in bifurcation analysis |

**Table 1** – Parameter values used in the fitting of current models, parametric analysis, and simulations of the full system, Eq. 13.

It has been already shown that Eq. 13 represents a minimal model where only $K^+$ conductances, without any $Na^+$ or $Ca^{2+}$ contributions could produce more complex dynamical behavior of $V_m$ than a stable fixed point $V_r$ [19].

Table 1. contains all the parameter values and/or value ranges used in the dynamical analysis of the model and the numerical simulations of transiently perturbed behavior.

## 3. Results

### 3.1 Parametric analysis of the $V_m$ dynamics

To translate the nature and intensity of the perturbations in different experimental preparations into changes of model parameters, we refer to (a) *static[3] alterations* of the astrocytic $K^+$ conductances for changes manifesting in the steady-state I-V relationships, and to (b) *dynamical perturbations* of $V_m$ in Eq. 13 for response to transient changes in the local field potential, depolarizations through the gap-junctions, or response to transient shifts of $[K^+]_o$.

Of not many disease modeling studies differentially isolating glial currents (mostly focusing on Kir), we looked for those where the changes *would* alter the monotonic I-V dependence of the total astrocytic current, Fig. 1. For the nature of current alterations, and the source references see Table A2 in *Appendix 2*. Sole focus on Kir currents with K2P current not being isolated has prevented those studies to report the impact of altered Kir current on the I-V curve of total current. The alterations of Kir currents summarized in Table A2 suggest that the model should consider asymmetrical changes of Kir conductivity in different $V_m$ ranges, either as a hallmark of a pathology or different functional channel expressions like in the case of heteromeric Kir4.1/Kir5.1 channels. Biophysical specificity of the Kir4.1/Kir5.1 channel is that it turns weakly rectifying Kir4.1 into a strong inward rectifier [45], numerically equivalent to almost fully attenuating the outward $I_{res}$ current.

Figure 6A illustrates the trend of change in the steady-state I-V curve of the total $K^+$ current in different scenarios of Kir deregulation in Table A2. A typical effect is a loss of monotonicity positive to $+60\ mV$, where both Kir4.1 and K2P currents coexist producing a non-monotonous N-shaped I-V curve. In Figure 6A we sketched this effect by different attenuations of Kir inward and/or outward component using our model, based on the I-V graphs in different disease models, Table A2. Included also, the lower solid curve, is the combined effect of loss of outward Kir conductance in parallel with drastically increased $[K^+]_o$ like in cases of seizures and spreading depolarizations. To produce the effects in Fig. 6A we were (a) decreasing $\bar{g}_{s-inw}$ for the inward range of WR-Kir current in the Hagiwara model, Eq. 3, and (b) doing the same using a multiplier in the GHK model, Eq. 8 for the outward range, due to a complex form of the conductance in $I_{res}$ model. The K2P current was kept in the nominal range and shape.

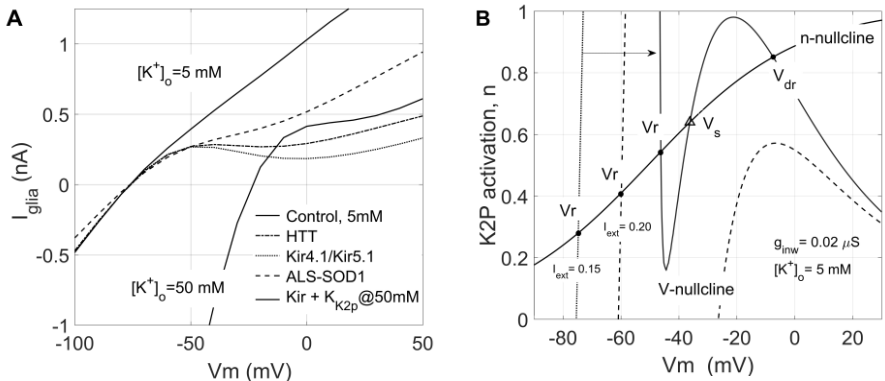

**Figure 6 – Loss of monotonic I-V curve via alterations of Kir4.1 currents in disease models.** (**A**) Alterations in Kir conductivity without changes in K2P conductance produce non-monotonic I-V curves. The lower solid curve illustrates the parallel effect of a drastic increase in external $K^+$, while all others are obtained by amplifying or attenuating the Kir currents in Eq. 13, at $[K^+]_o = 5mM$ based on the nature of deregulation in Table A2, Appendix 2. The labels in the legend relate each curve to the corresponding study in Table A2. (**B**) Example of nullcline analysis in $(V_m, n)$ plane. Increasing the constant $I_{ext}$ changes (into an N-shape) and shifts the V-nullcline introducing a saddle fixed point $V_S$ (triangle), and markedly more positive stable node or focus $V_{dr}$ via the fold bifurcation. The dotted, dashed, and the solid line show the changing shape of the V-nullcline with $I_{iext}$.

Since the typical dynamic perturbations, same as with neurons, transiently depolarize the glial membrane by $I_{gjc}$ or $I_{lfp}(t)$ currents, the principal bifurcation parameter was a constant in time $I_{ext}$.

---

[3] Even though the experiments observe real-time channel expressions changes, those happen on time scales of minutes, hours or longer, which warrants keeping the corresponding parameters constant on timescales of milliseconds or seconds.

As a second bifurcation parameter, we varied $\bar{g}_{s-inw}$ and $[K^+]_o$ with attenuation of $I_{res}$ as a static perturbation of the model, so that all scenarios of impact on Kir conductivity could be implemented.

In the 2-dimensional system, Eq. 13, using nullclines let us detect the trends of change in the phase portrait. As expected and previously illustrated in models of neurons very similar to Eq. 13 [19], chapter 5, in addition to the nominal equilibrium state $V_r$ - a stable node for $V_m$ slightly more positive to $E_K$, the N-shaped I-V curve introduces a second stable node via the fold bifurcation, as we vary $I_{ext}$, Fig. 6B. The fold bifurcation is generic in 2-d models like Eq. 13 with N-shaped nonlinearity in the voltage[4] equation and robustly observed within wide parameter ranges. We keep referring in the rest of the text to the nominal stable, resting state as $V_r$, and to the more depolarized, again stable steady-state introduced by the fold bifurcation as $V_{dr}$.

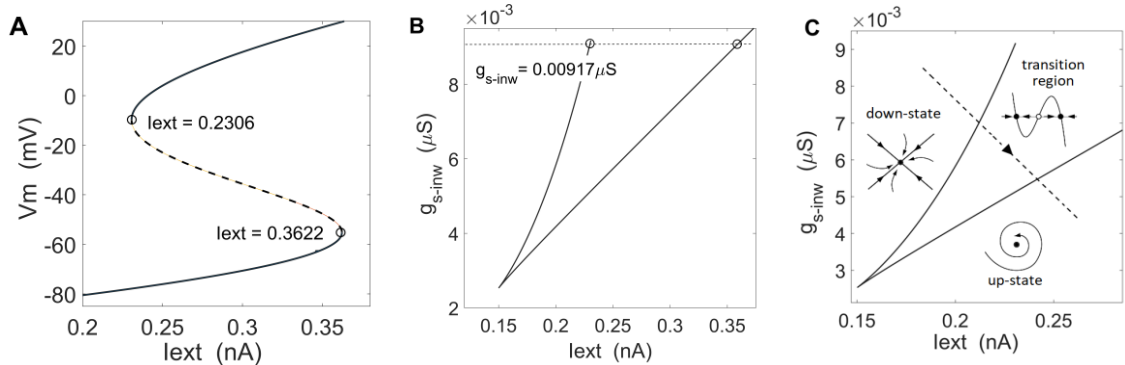

**Figure 7** – **Fold bifurcations in $(V_m, n_{K2P})$ dynamics** are generic when the perturbed model displays an N-shaped I-V characteristic. (**A**) Fold bifurcation diagram of Eq. 13 for $g_{s-inw} = 0.00917\ \mu S$, with $I_{res}$= 0. (**B**) A cusp curve in $(I_{ext}, g_{s-inw})$ plane obtained by running 2-parameter continuation from the fold point at $I_{ext} = 0.2306$ in (A). The coordinates of the corresponding stable steady states are $(V_r = -77.86\ mV, n_{K2P} = 0.0964)$, and of the saddle-node close to $(V_{dr} = -9.734\ mV, n_{K2P} = 0.6)$, with $g_{s-inw}$ kept at the same value as in (A). (**C**) The dashed line shows a typical path through the parameter plane corresponding to a membrane with deregulated Kir conductance perturbed by a net depolarizing current $I_{ext}$. The region bounded by the cusp curve separates the two *monostable* parameter domains, the down-state domain, corresponding to the true nominal resting state $V_r$ (of node type), and that of the up-state, corresponding to a *single* depolarized steady state $V_{dr}$, of focus/spiral type. For parameter values inside the transition region, within the cusp, the model displays two stable nodes separated by an unstable saddle point.

Figure 7A and 7B show the fold and cusp bifurcation diagrams for changing $I_{ext}$ and $g_{s-inw}$, with the outward $I_{res}$ attenuated. All continuations were done using the *AUTO* package [46] as implemented within *XPPAUT* [47]. Examples in Fig.7A and 7B were computed with $I_{res} = 0$, though qualitatively the same behavior is present with $I_{res}$ between zero and 15% of its intensity. All other model parameters were kept at the values given in Table 1, used in the fitting of steady-state I-V characteristics, with $[K^+]_o = 2.5\ mM$ and the glial leak $g_{leak} = 0.0013\ \mu S$. At $I_{ext} = 0.2306\ nA$, in Fig. 7A the bifurcation introduces a saddle-node at $V_{dr} = -9.734\ mV$, notably depolarized to the nominal $V_r = -77.86\ mV$, which further splits into a saddle and stable node at $V_{dr}$ as $I_{ext}$ further increases. The cusp curve Fig. 7B fusing fold bifurcation behavior of two parameters divides the $(I_{ext}, g_{s-inw})$ plane in three regions: (I) the *down-state* represented by $V_r$, (II) the *up-state* represented by the depolarized resting voltage $V_{dr}$ of type focus, and (III) the bistability region where $V_m$ transits towards either of the stable nodes, qualifying the model for switching behavior. Equations 13 represent the first minimal ODE model of the glial membrane near $V_r$, based on whole-cell recordings, suggesting the existence of a stable, depolarized up-state in glial $V_m$ dynamics, within a wide range of parameters.

The biological relevance of downregulating or abolishing the WR-Kir outward $I_{res}$ current as a parametric perturbation is directly suggested by (i) electrophysiological data in several disease models we listed [48], [49], as well as (ii) by the specific properties of heteromeric Kir4.1/Kir5.1 channel populations [50], [45]. The later accounts for a nonnegligible fraction of Kir4.1/Kir5.1 channels in astrocytes in different brain regions [51], [52], [12]. The extent of deregulation of the

---

[4] We refer to the first ODE in Eq. 13 as *voltage equation* because it describes the $dV_m/dt$ derivative, even though as a physical law it is the Kirchhoff's current equation.

*inward conductance of the Kir4.1 alone* – a decreased slope in the I-V curve negative to $V_r$, is not detrimental for the observed qualitative changes since it doesn't produce by itself the N-shaped I-V curve.

The dashed line in Fig. 7C represents a typical *perturbation line* of Eq. 13 combining (i) permanent deregulation of outward conductance of WR-Kir, with (ii) changing inward WR-Kir conductance in experimental conditions where an astrocyte is (iii) subjected to depolarizing current input $I_{ext}$.

Slow depolarizations induced by $K^+$ accumulation in extracellular space (ECS), and subsequently the positive shift of $E_K$ and $V_r$ suggest that either as a parameter, a transient perturbation, or a slow variable in an extended 3-dimensional model, $[K^+]_o$ could be inducing instability of $V_r$. Since an extension of the system, Eq. 13 incorporating $[K^+]_o$ dynamics in $R^3$ or higher [53] is beyond the scope of this study, we verified that fold and cusp are robust in $(I_{ext}, [K^+]_o)$ parameter domains as well, with $I_{ext}$ in the same range as above, and $[K^+]_o$ modestly raised compared to the nominal 2.5 $mM$, Fig. 8. It is important to report that higher the $[K^+]_o$ much less pronounced attenuation of $I_{res}$ produces the N-shaped curve and the fold bifurcation, (much less than 85%).

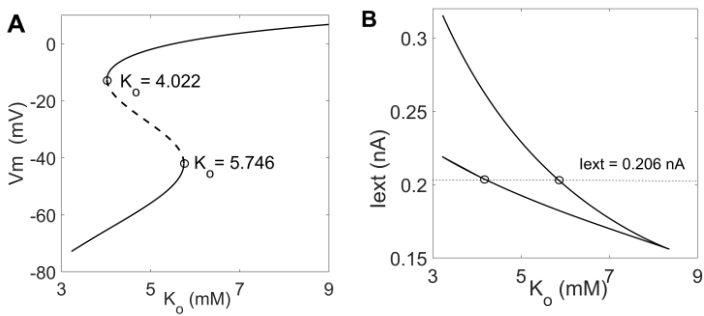

Figure 8 - **Fold bifurcations under variable $[K^+]_o$ with abolished $I_{res}$** (A) Fold bifurcation in Eq. 13 for variable $[K^+]_o$, keeping $g_{s-inw} = 0.00951\ \mu S$. (B) A cusp curve in $(I_{ext}, [K^+]_o)$ plane obtained by running 2-parameter continuation from the fold point at $[K^+]_o = 4.022$ in (A). The actual continuations in XPPAUT were computed using $E_K$ due to better numerical convergence. $[K^+]_o = 4.022\ mM$ corresponds to $E_K = -89.32\ mV$, for $[K^+]_i = 130\ mM$.

Large separation of time scales between $V_m$ responses and $[K^+]_o$ transients, from milliseconds to second and minutes, warrants analyzing the basic $[K^+]_o$ impacts on $V_r$ stability treating it as a parameter in the simple 2-dimensional system, Eq. 13.

Quantitative studies summarizing the properties of macroscopic currents through the glial Kir4.1 [54] and K2P-TREK channels [28], [37] suggest that apart from K2P activation there is no other channel kinetics critical to extend the minimal model, Eq. 13, with additional dynamical variable. Whole-cell Kir4.1 responses to voltage clamp commands suggested almost instantaneous activation with rising times of less than 2 ms. The instantaneous Kir4.1 currents, measured at 2 ms, were somewhat higher in the mid and outward range but not producing qualitative change in I-V curve. There is therefore no candidate for a *resonant variable* in the 2-D model, Eq. 13, [19], [55] which would induce a cyclic behavior in $\mathbb{R}^2$. A word of caution is needed in this context due to the measurement difficulties with the leaky astrocytic membrane patches. Inability to precisely determine the input and access resistance of the patched astrocyte [1] makes it unreliable to compensate for the serial loss, which in turn always produces underestimated maxima of both, instantaneous and steady currents potentially compromising our judgment on the rise times in the current responses to voltage commands.

### 3.2   Numerical simulations of the bistability and switching

For demonstrating the switching capability we perturbed the model, Eq. 13, using time series from recordings of glial depolarization during *electrographic seizures* induced by transient elevation of $[K^+]_o$ [9]. The glial $V_m$ was recorded using two-electrode *in-situ* protocol on glial membrane in rodent hippocampal slices.

Figure 9, (upper trace) shows a sample of glial depolarization $\Delta V_{ext}$ in response to neuronal seizure, Fig. 11A in [9], digitized from a printed image using the *Graph Grabber* tool [56]. Such signal features: (i) a fast depolarizing, tonic transient of $\Delta V_m \approx 20 mV$ within initial ten seconds, followed by (ii) pseudo-periodic clonic episodes lasting in total for nearly 60 seconds. To use the $\Delta V_{lfp}$ recording as a perturbing signal in our model we needed somewhat realistic time course of the corresponding

$\Delta E_K(t)$ shift, for which purpose we used $\Delta V_{ext-dc}$ averaged signal of 27 seizure events (from 5 cells) from the same study [9] see Fig. 10A, plotted as a lower curve in Figure 9 below, so that for a realistic example of $I_{ext}$ in Eq. 14 we get:

$$I_{ext}(t) = I_{gjc} + I_{lfp}(t) = I_{gjc} + g_{lfp}\Delta V_{ext}(t) \quad (nA)$$
$$E_K(t) = E_K + \Delta E_K(t) \approx E_K + \Delta V_{ext-dc}(t) \quad (mV) \quad (15)$$
$$\Delta V_{ext}(t) = V_{glia}(t) - V_{lfp}(t); \quad \Delta V_{ext-dc} = \langle\Delta V_{ext}\rangle - \langle\Delta V_{lfp}\rangle \quad (mV).$$

The $V_{ext}(t)$ signals were measured with a reference to a distant ground in the bath and should be referenced to the LFP level immediately adjacent to the cell measured the same way [9]. That correction averaged at 3 mV, was omitted in the final simulations since it produced no difference and would in principle cancel out in $(V_{ext} - E_K)$ if such control measurement of externally generated driving force existed.

Perturbation signals as in Eq. 15 mimic a transient depolarization event of a hippocampal glial cell *in situ* - depolarized by a positive, outward current resulting from sustained seizure-like discharges invading the surrounding neurons in a larger region, slowly and transiently elevating, and restoring the nominal $[K^+]_o$ during 72.6 seconds. We do not use it to simulate bath application of a constant elevated $[K^+]_o$ in [9], but try to simulate more realistic transient depolarization which in reality comes with transiently altered local $\Delta E_K(t)$.

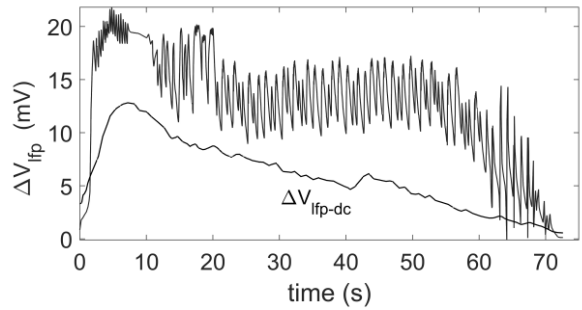

**Figure 9 – Transient glial depolarization signals –** Depolarization transient $\Delta V_{lfp}$ from resting (upper trace) measured on glial soma *in situ*, in response to an electrographic seizure initiated by increased $[K^+]_o$ to 8.5 mM, in a rat hippocampal slice. A two-electrode protocol was used with a reference electrode in the bath. $\Delta V_{lfp-dc}$ (lower trace) is an average of 135 depolarizing episodes, recorded from 5 cells, used to approximate the shift of the Nernstian potential $\Delta E_K(t)$ due to $K^+$ accumulation.

We find $\Delta V_{ext-dc}$ to be a good example of $\Delta E_K(t)$ transient assuming it represents an average of LFP events that (a) in reality are subjected to low-pass filtering of the ECS [57] which will reduce them to the slower timescales typical for gradual $[K^+]_o$ changes, and (b) are of common shape in the initial steady depolarization phase, time-aligned at the onset. Its maximal amplitude of $12.8\ mV$ is close to the Nernstian $\Delta E_K = 13.7\ mV$, for a relative shift from $[K^+]_o$ at the nominal 5 mM in our data (the nominal I-V relations in the model), to 8.5 mM bath concentration robustly producing the seizures [9]. The slower rise time in the initial 10 seconds, as well as the steady decay of the averaged $\Delta V_{ext-dc}(t)$ probably reflect the joint action of the Na-K pump and other restorative mechanisms of ion homeostasis. Such $\Delta E_K(t)$ is therefore a safe approximation for low to moderate $[K^+]_o$ increases on the timescale of seconds, even a more conservative one, producing slightly less depolarized $E_K$ shifts if we start from a baseline of $2.5\ mM$ of external $K^+$. Unspecific conductance $g_{lfp}$ was introduced to convert the transient voltage perturbation $\Delta V_{ext}(t)$ into current perturbation signal in Eq. 15. Indicative value range of $g_{lfp}$ was set between $0.01 \div 0.016\ \mu S$ to keep the total $I_{ext}(t)$ in simulations within the range of the fold bifurcation diagram, Fig. 7A, matching the dynamic range of the clonic discharges in $\Delta V_{ext}$ of $5 \div 10\ mV$, from $t = 10\ s$ onwards, Fig. 9.

A simulated example of switching behavior in response to the above perturbing transients is illustrated in Figure 10. We observed two qualitative behaviors depending on $I_{res}$ attenuation and $I_{ext}$ and $g_{s-inw}$ value ranges. With fully abolished outward $I_{res}$, constant $I_{gjc} = 0.125\ nA$ and nominal $g_{s-inw} = 0.00917\ \mu S$ at $[K^+]_o = 2.5\ mM$, Fig. 10A shows a series of *plateau responses* on $\Delta V_{ext}$ from Fig. 9, with $0.01 \leq g_{lfp} \leq 0.017$. With increasing $g_{lfp}$ the response switched from the downstate to the upstate in a threshold manner – from partially to fully. With $I_{res}$ attenuated to 15% of its amplitude, a more *erratic switching* profile without prolonged upstate was a typical response, shown in Fig. 10B with the other parameters kept the same.

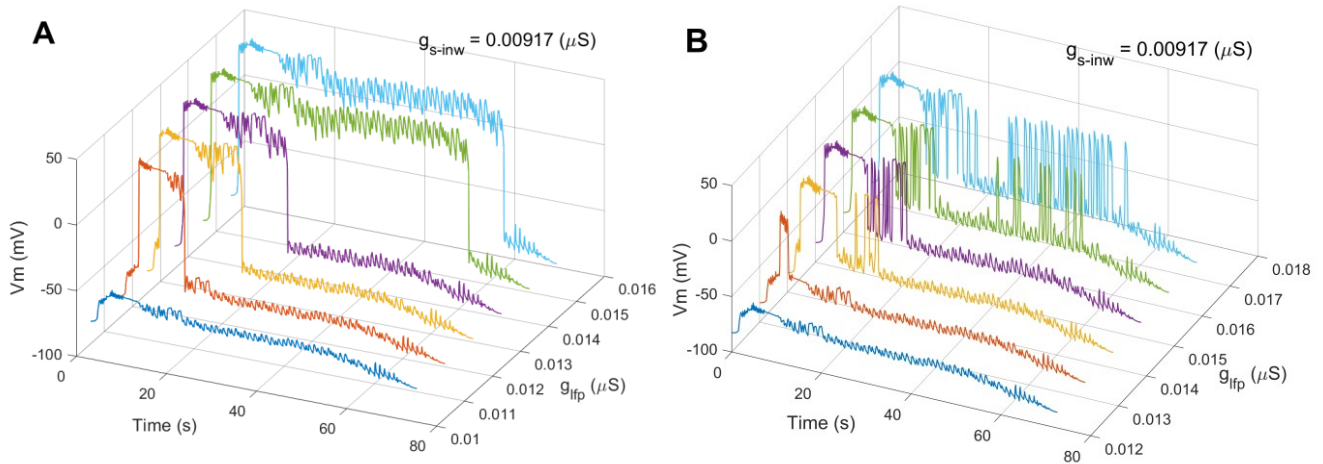

**Figure 10 – Glial switching on transient seizure-like perturbation** – (**A**) Simulations of the full model $V_m$ response on electrographic seizure, Fig. 9, with fully abolished Kir outward $I_{res}$ current and increasing $g_{lfp}$ show plateau switching between downstate and upstate, corresponding to $V_r$ and $V_{dr}$. (**B**) The same simulation, with $I_{res}$ amplitude reduced to 15%, demonstrating an irregular, erratic switching with a frequency following that of the clonic phase of seizure signal. The presence of $I_{res}$ shifts both $V_r$ and $V_{dr}$ slightly more negative. A constant level of external current through gap junctions was kept at $I_{gjc} = 0.125\ nA$ and the slope of Kir inward conductance was held at the nominal $g_{s-inw} = 0.00917\ \mu S$.

Decreasing Kir inward current, via $g_{s-inw}$ led to more pronounced instability and more frequent switching in either profile for lower transfer conductance $g_{lfp}$, see Figure S1 in *Supplementary Information*. No particular, numerical sensitivity in any of the parameters was observed indicating a more peculiar singular behavior, in dynamical terms.

## 4. Discussion / **I** ~ *importance*, **S** – *significance*, **N** – *novelty*, **U** - *utility* /

The observed dynamical instability could point to different areas of impact depending on the physical model system and the specific $V_m$ dependence of the modeled phenomenon. A general conclusion in hypothesis-free terms would be that with its very basic ion channel composition the glial membrane responds differently and distinctly to external depolarizations when Kir or combinations of Kir and K2P conductances are deregulated, due to the interplay of Ohmic and complex non-Ohmic conductances, nonlinear in $V_m$. Below, we discuss the possible implications of the presented biophysics from several different neurobiological perspectives.

### 4.1 On the model

**(I1)** Numerous biological *association studies* have related, *put in context* the changes in expression of Kir4.1 currents and the biophysical properties of measured currents to the altered function of glia. In translating such changes into specific roles of the cell, it has remained elusive which mechanisms could transduce the molecular changes into a whole-cell voltage response. We modeled $V_m$ dynamic of an isolated astrocyte by a minimal ODE model incorporating the major $K^+$ conductances, WR-Kir and K2P, and a leak current to explore the direct effect of nonlinear changes in the total astrocytic ion conductance to $V_m$ dynamics. In other words, it models the simplest form of modulation of $V_m$ by alterations of ion conductances with no other signal transductions being modeled, like membrane mechano-sensitivity, pH sensitivity, actions of physiological agonists, etc. The model, Eq. 13, let us ask the typical question - which nature of (de)regulation of the conductances impacts the stability of the nominal resting state $V_r$, which in turn controls all electrogenic transport mechanisms.

**(I2)** Decades of conductance-based models suggest that even very small currents need careful consideration since the nonlinear interplay of their conductances could impact the stability of the resting state and in turn the cell response [19]. That has motivated the incorporation of the K2P current through its activation dynamics, $n(t)$, as a second dynamical variable.

The major perturbations of the baseline model came from decreased Kir conductance in the outward $V_m$ range, with K2P conductance remaining unchanged or eventually increased, resulting in a non-monotonic steady-state I-V curve. **(N1)** In that regard we present the first ODE model of glial $V_m$, Eq. 13 which incorporates two distinct and *differentially regulated* $K^+$ currents, in addition to the unspecific $K^+$ leak current, based on recordings from isolated cells.

**(S1)** Non-monotonous, N-shaped I-V relation produced by a prominent decrease of WR-Kir in the outward range, gives birth to a new, more depolarized *stable* steady state $V_{dr}$ for physiological average external depolarizing current, $I_{ext}$. When present, such a mathematical feature of a model of a non-excitable cell is generic and observed robustly in a wider parameter range, Fig. 7. We argue here that the emergence of a distinct and stable $V_{dr}$ introduces a specific and distinctive transition - translating the loss of pseudo-Ohmic, linear behavior of the membrane into a *bistable switch* when exposed to depolarizing transients, Fig. 9. Bistabile switches with or without hysteresis are pervasive and diverse in biological systems since those are the simplest outcomes of instabilities of the resting state [58]. An elegant analytical model of the same form of bistability in hypokalemic skeletal muscle is presented in [59].

To simulate the switching between $V_r$ and $V_{dr}$ using a perturbation specific to the repertoire of real glial cells we used a seizure-like transient recorded on glia rather than a ramp or constant $I_{ext}$, since glial cells in real circuits do not receive prolonged depolarizations by constant external currents. We believe the chosen electrographic seizure has a typical waveform for a wider range of strong transient depolarizations resulting from coherent discharges of hippocampal neurons.

In modeling open, background Kir channels we departed from the typical modeling approach, a single current – single biophysical description. The reversal of polarity of the driving force changes the nature of the physiological block resulting in *qualitatively different pseudo-gating in the WR-Kir pore for inward and outward flux* (Appendix A). This assumption warrants the use two different biophysical descriptions for additive inward and outward fluxes. Figure 3 suggests that this conceptual approach results in an overlapping $V_m$ regions where both mechanisms (blue and red curve in Fig.3) produce a small outward current. Those outward permeation events happen for very small $\Delta V_m$ where the electrical enthalpic contributions are comparable to the fluctuations – resulting in outward pseudo-gating in the central cavity, as well as within the SF, described respectively by the Hagiwara model, and by the nonconstant-field GHK equation, Eq. 5. Such different gating scenarios require several effective charge valences $z_{inw}$, $z_B$, and $z$ to appear within descriptions of WR-Kir current. The equivalence of the form of nonconstant-field GHK model of $I_{res}$, Eq. 8, and the basic form of rate model based on simplified Eyring's formalism with a single, centered, and symmetrical barrier [42], together with the attempts to relate Hagiwara's model to multi-barrier Eyring model [60] suggest at least an opportunity to consider bringing the proposed separate models of WR-Kir inward and outward fluxes under a more coherent formalism.

**(N2)** Departing from the constant field assumption in the Nernst-Planck formalism of an open pore, results in a modified Goldman-Hodgkin-Katz current equation which allows incorporating simple forms of nonlinear energy profiles within the pore, as long as the integral in the current equation is solvable [27]. We applied it for the first time to describe the outward component of WR-Kir current, Eq. 8, and by extending it with quadratic dependence of the barrier peak, Eq. 7, the observed current saturation with voltage was captured by the model, Fig. 4. Accumulating evidence of complex gating in open pores, lacking a voltage-sensitive gate, may qualify this approach in a wide range of leaky channels.

Preliminary extension of Eq. 13 we attempted with $[K^+]_o$ dynamics [53], [61], [62] *in a very simplified, spatially constrained ECS* compartment demonstrated an enriched dynamical repertoire of the model and will be reported in another study. Within $[K^+]_o$ ranges indicating the richer bifurcation repertoire in [53] and [63], we observed some entrainment of the glial $V_m$ in Eq. 13 by the $[K^+]_o$ dynamics driven by a coupled neuronal bursting [53], but those require more precise dynamical characterization. Other important extensions of the model should include (a) incorporating the chloride conductances explicitly and (b) adding mechanosensitivity to K2P conductances to simulate the osmolar stress of the membrane.

Very limited availability of transient recordings from glia leaves us without traces of real-time perturbations of glial $V_m$ on different time scales which is a limiting factor to considering a *non-autonomous, driven ODE* in the model Eq. 13, since in immediate proximity of spiking neurons the glial cells are never in a true voltage steady state. An isolated astrocyte is not a biological reality but has been an experimental and modeling target system for detecting and describing the baseline cell properties. The heavily interconnected glial cells in the *glial cell continuum* require testing any indications of $V_m$ multistability using spatially extended models in the light of experimentally observed $V_m$ *isopotentiality* [64] – a tendency of interconnected glia to quench any marked depolarization.

## 4.2  On the instability and glia

**(S2), (N3)** Translating $V_m$ bistability and altered WR-Kir current into effects significant for the whole cell function, the following impacts are directly implicated by the model:

- **Depolarized resting state $V_{dr}$ as a *catastrophic event*[5]** – Even though it is one of the simplest instability scenarios, the birth of a stable, very depolarized resting state is potentially *catastrophic* since switching to $V_{dr} > -10\ mV$ is taking the membrane and the cell, to a state with much higher potential energy than the nominal hyperpolarized resting state, not corresponding to the actual $[K^+]_o$. Prolonged depolarization of the glial membrane around $V_{dr}$ would impact all glial functions, from metabolic support to membrane transport, and practically all neuromodulatory roles assigned to glia.

- **(I3) Voltage switching between $V_r$ and $V_{dr}$ disrupts the kinetics of active transport mechanisms** - Since all major electrogenic pumps and transporters show some extent of voltage dependence [68], [69], [70], the depolarized $V_{dr}$, or the up-state should violate the assumptions made for the steady-state transport rates near $V_r$. In the case of ATP-driven pumps, in the Na-K pump, for example, all simplified descriptions used in computational models, in a form of a simple current generator $I_{pump} = \rho f([K^+]_o, [Na^+]_i)$ focusing on the "acceleration" term $f$ which models the sensitivity to ion concentration changes, are warranted by the existence of a constant steady-state rate $\rho$, assumed to be voltage-independent. When $V_m$ switches to $V_{dr} > -10\ mV$, dramatically shifted $\Delta V_m$ to large positive values impacts some of the binding and translocation steps in the kinetic model of the pump and questions the biophysical plausibility of keeping $\rho$ constant. In particular, where the perturbing voltage transients like the electrographic seizure we used, produce a jump near $V_{dr}$ *not resulting* from moderately elevated $[K^+]_o$, i.e., $E_K$ not much above $V_r$, a large and lasting positive transient is produced. If further studies prove the bistability in $V_m$ is electrophysiologically plausible, moving to a more detailed macroscopic model of the pump might be required with $\rho$ explicitly depending on the full form of the electrochemical potential. For the pump rate $\rho$ in units of molar flux, we can write [69]:

$$\rho \sim \Delta G + \Delta \tilde{\mu}_H\ (mM/s), \quad \text{where } \Delta \tilde{\mu}_H = \Delta \mu + zFV_m\ (RT\ units),$$
$$RT \ln(c_{ext}/c_{in}) + zFV_m < -\Delta G,\ \text{for active transport,} \tag{16}$$

where $\Delta G$ represents the purely chemical potential energy from ATP hydrolysis, and within the electrochemical part $\Delta \tilde{\mu}_H$ we recognize the Nernstian, or osmolar contribution $\Delta \mu = RT \ln(c_{ext}/c_{in})$ and the electrical part $zFV_m$ that impacts $\rho$ under prolonged depolarization at $V_{dr}$. An implicit assumption not typically discussed is that only near the nominal equilibrium $V_r$ where the electrochemical gradient $\Delta V_m$ is small and the osmolar contribution large and steady, the $FV_m$ term could be eventually neglected. The inequality in Eq. 16 is the typical condition for a sustained primary active transport, with sign convention in $-\Delta G$ representing the amount of free energy available from hydrolysis of a mole ATP, rather than the barrier height in the enzymatic reaction [71], [69]. Some variants of the macroscopic model of the pump have kept the voltage dependence [72] and have produced effects in whole-cell dynamics attributable to the Na-K pump. In such models, the eventual effects of voltage bistability could be directly tested. Stoichiometry coefficients have been omitted in Eq. 16 for simplicity.

---

[5] We use the term in the context of the *catastrophe theory* of dynamical systems [65], [66] and *catastrophic bifurcations* [67], initially formulated over the impact of the same instability we observed - the fold bifurcation, on the global behavior of both, natural and man-made systems.

The electrogenic sodium bicarbonate $Na^+/HCO_3^-$ cotransporter (NBCe1) specifically implicated in glial metabolic and other functions, is another important candidate to be analyzed in the context of bistability since the reversal of the polarity of $\Delta V_m$ would cause switching of the transport direction and in turn the pH modulation of the cell between alkalization and acidification, and consequently of the immediate extracellular environment [7].

Of the other transporters potentially "vulnerable" to bistability let us mention the astrocytic GABA transporters (GAT1-GAT3) expressed in local cortical circuits [73], [74], modeled macroscopically as explicitly $V_m$ dependent [75].

- **(I4),(S3) Switching between the two resting states may introduce erratic transients in trans-junctional voltage over the GJCs –** It has been shown that the conductance and rectification of GJCs are dependent on the trans-junctional voltage $V_J$ between the connected cells across a wide range of the Connexin isoforms [76], [77], [78], and specifically for the Connexin43 hemichannels, one of the two most abundant subtypes assembling the astrocytic GJCs, with Connexin30 [79]. Due to the nonlinear steady-state I-V relationship and the nonlinear $g_{gjc}(V_J)$ in a wider $V_J$ range in most GJCs [80], [81], the complex two-stage gating of the hemichannels may result in erratic behavior of the GJC if the connected cells are prone to switching between markedly separated $V_r$ and $V_{dr}$. The notable difference between the *instantaneous and steady* $g_{gjc}(V_J)$ [77], introdcues another major uncertainty in the eventual models [81]. The origin of such uncertainty between the residual and fully conductive conformations could be in principle justified by an atomic level model of GJC electrostatics [82] suggesting complex Coulombic profiles, potentially vulnerable to erratic voltage transients.

- **(I5), (S4) Glial encoding of $[K^+]_o$ variations as a signal gets a distinct state** – In the present understanding of the *encoding capability* of glial membrane, the glial $V_m$ with certain low-pass filtering capability (with $\tau_m = C_{eff}/g_{slope} \approx 10ms$ for 20pF patch capacitance) close to linearly encodes $[K^+]_o$ variations as a signal through $(V_m - E_K)$. We can also say $E_K$ itself smooth out the real-time $K^+$ variations from neuronal discharging partly due to the low-pass filtering of the ECS to volume transmission signals [57] and partly due to the $\log(c_{out}/c_{in})$ nature of Nernstian dependence. The emergence of $V_{dr}$ introduces therefore a distinct, depolarized state potentially detectable on various timescales by a suitable measurement setup. In other words, $V_r$ and $V_{dr}$ encode (a) an almost linear potassium electrode, and (b) a deformed, N-shaped I-V relationship of deregulated conductance(s), respectively, Fig. 6A.

The presumed ion-homeostatic role of Kir4.1 in glia was not discussed in the context of bistability because the differential contributions of the different ion transport mechanisms are still elusive [83], and a reasonably reduced mathematical model addressing the spatial complexity of volume transmission in neuroglial circuits requires a dedicated experimental study for detecting the ranges of those parameters critical to the ion homeostatic effects.

In summary, the proposed minimal 2-dimensional model of the interplay between WR-Kir and K2P currents in astrocytic membrane shows that deregulations turning WR-Kir current into inwardly rectified, robustly produce a second very depolarized steady state of $V_m$ and a whole-cell behavior considered markedly non-physiological within glial biology. To prove or refute such bistability as neurobiologically plausible, a measurement protocol well suited for transient $V_m$ changes would be needed, probably employing voltage imaging, a dynamic (current) clamp, and cell assays displaying deregulated Kir4.1 or sufficient densities of Kir4.1/Kir5.1 channels of glial origin.

**(U1)** The proposed model, Eq. 13, is of general biophysical utility, directly applicable to the interplay between Kir and K2P currents in membranes of other systems where K2P channel variants display nonnegligible activation kinetics, like in the heart and kidneys, see [18] for more cases. Bistability showing comparable voltage separation of $V_r$ and $V_{dr}$ like presented has been already observed in cardiomyocytes in hypokalemic conditions [14], where strongly-rectifying Kir2.1 channels interwork with the linear K2P-TWIK channels.

**Author contributions** – PJ and LK designed the study. PJ formulated the models, did the parameter analysis, and wrote the paper. PJ and DS did the numerical simulations. PJ and LK assessed the results.

**Declaration of Interests** - The authors declare no competing interests.

**Acknowledgments** – We thank Prof. Christian Steinhäuser and Dr. Gerald Seifert for sharing with us the unique recordings from isolated astrocytes [12] of both Kir and K2P currents, for their lasting support in getting a sense of data and experimental limitations, as well as to CS on the numerous discussions and critical inputs. We thank Prof. Stephen Traynelis for warranting us to digitize and use the printed graphs from [9]. We thank Prof. Meyer Jackson for his comments on the model of $I_{res}$ current, and for his inspiring treatment of biophysics of open pores [27] which motivated our modeling on several points. We thank Prof. Pavle Andjus for his comments on the simulation results. This study was partly funded by the National Institute of Mental Health (US), grant numbers MH117769 and MH125030 to PJ.

The XPPAUT .ode files for the bifurcation analysis and simulations will be available in *ModelDB* at https://senselab.med.yale.edu/ModelDB, under the model name "*Glial voltage dynamics driven by Kir & K2P currents*".

## Appendix-1   Short-pore model of WR-Kir channel and physiological $Mg^{2+}$ block

We base our model of Kir current on (i) $K^+$ conduction through a *short pore* (Fig. A1), common to all $K^+$ channels - confined to the part of the transmembrane structure, consisting of a selectivity filter (SF) extending into the internal water-filled *central cavity* [84], [85] and on (ii) *physiological blocking* of the pore by cytoplasmic cations as a permeability modulation mechanism. Coulombic interactions within the short pore are therefore used to explain the permeation mechanisms as an interaction between the blocking $Mg^{2+}$ and permeant $K^+$ ions, as a basis for a weak rectification. Unlike in strong inward rectification, where both, the internal $Mg^{2+}$ and/or polyamines fully block outward flux [34], [86], in the outward permeation scenario of a weakly rectifying pore (Fig. A1C) the partial and flickering block allows the entrance of permeant ions into the SF and modulates their electrodiffusion at high rates. This pore model is crucial for explaining the *anomalous*, inward rectification of Kir conductances, which can't be described by Nernst-Planck (NP) formalism based on electrochemical gradients over open, leaky channels. The short pore model applied to outward conduction in WR-Kir channels, allows extending the NP model with a non-constant field and in turn a voltage-dependent permeability, equation (9), under the assumptions of a changing Coulombic profile with voltage. These extensions formulate the main assumptions of the $I_{res}$ model [27].

In line with the above assumptions, recent evidence from single-channel recordings and molecular dynamics simulations of the potassium KcsA pore [87] suggests that the voltage and ligands induce *structural fluctuation* changes within the electrostatics, Coulombic profile of SF which produces a universal pseudo-gating of an open pore at SF, *without* introducing a conformational change and gating.

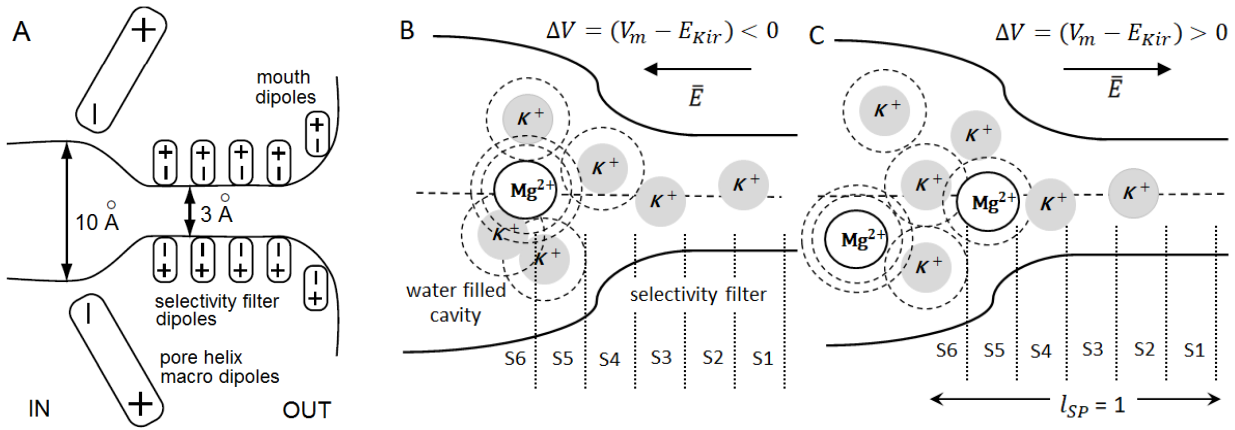

**Figure A1 - *Short pore* model of permeation through generic weakly-rectifying Kir channels** – **(A)** A *short-pore* view of the Kir pore consisting of a selectivity filter (SF) and the inner, water-filled *central cavity*. The sketch shows the sources of electrostatic interaction along with the transmembrane structure acting on permeant and blocking ions, modified from [88]. Inside the SF, selectivity filter dipoles create binding sites (S1 to S4), corresponding to energy wells where permeant ions are stabilized between the transitions. **(B)** and **(C)**, A simplified one-dimensional view of the short pore showing the different voltage dependence of the $Mg^{2+}$ block and the Coulombic interaction with the permeant ions. From negative to positive $\Delta V_m = (V_m - E_{Kir})$, and constant or slowly varying $[K^+]_o$, the resulting force on $Mg^{2+}$ and $K^+$ ions changes and the resulting blocking events are qualitatively different, see text for details. Arrows show the direction of the electrical field $\bar{E}$ on a positive ion along the reaction coordinate. We assign a unitary length $l_{SP}$ (shown in C) to the extent of the pore over which we model potential energy landscape. Dashed circles around the ions in (B) and (C) represent the full or partial hydration shells. The length of the SF without the outer vestibule in (B) and (C) is disproportionally shorter than in (A), to keep the image compact.

The short-pore model assumes the concentration of $Mg^{2+}$ and $K^+$ within the central cavity and at the external pore vestibule, correspond to the physiological intracellular and extracellular bath concentrations respectively. Local, *relative, nonequilibrium concentration changes* due to effects like the attraction of cations by rings of negatively charged residues, in vestibules on either side of the SF entrance, are neglected.

To describe both, the inward and outward permeation, we assume that the transmembrane voltage $V_m$ drops over the whole *short pore* - over the whole cavity and the SF, but the fraction $\delta$ introduced in Eq. 7 reduces it to $\delta \Delta V_m$ influencing with a smaller fraction the middle of the pore and S5 position where the outward barrier is centered, as marked in Fig. 3B. For completeness, the S0 position (not shown in Fig. A1B and A1C) would be just outside the SF, in the outer vestibule. This is a similar approach to the one used in permeation studies of weak rectification in renal Kir1.1 / ROMK1 [89, 90]. This pore model suggests that the blocking cation moves between two *modulatory* sites [91] depending on $\Delta V_m$ polarity, thereby defining the main rectification barrier(s).

*Inward pseudo gating* - For negative and very small positive net driving forces $\Delta V_m$, (Fig. A1B), the resulting electrostatics from pore helix macro dipoles keeps blocking $Mg^{2+}$ ions within the cavity (designated by position S6), so that the inward rectification results from a competition of permeant $K^+$ and $Mg^{2+}$ ions for the central, axial positions within the cavity. Even though $\Delta V_m$ influences S6 position with a much smaller fraction, not exceeding $\delta \Delta V_m$ (Fig. 3) the dependence of inward flux on $V_m$ for $\Delta V_m < 0$ is still exerted on the $K^+$ ion in S6 through the multi-ion file within SF (binding positions S1 to S4), which contributes to the resulting Coulombic forces and ion-crowding within the central cavity (represented by S5 and S6 positions). By resulting force, in this simplified view, we mean Coulombic repulsion between the permeant ions acting in parallel to the stabilizing attractive force on them coming from the negative end of pore helix macro dipoles (Fig. A1A), tilted so to make the center of the cavity electrostatically favorable point for a cation [26]. Hagiwara's model of the macroscopic inwardly rectified Kir currents assumes the existence of equilibrium pore-open probabilities of inward permeation suggesting a Boltzmann term in Eq. 3 for describing the macroscopic conductance (rather than barrier crossing rates), analogous to the open and closed probabilities in gated ion channels, see [34] for Kir-specific review.

***Outward pseudo gating*** - In the reverse case, for a more positive driving force $V_m > -50 mV$ (Fig. A1C), an outwardly directed net driving force acts on permeant and blocking cations, pushing partly or fully dehydrated $Mg^{2+}$ ion towards the entry of the SF, which results in a partial flickering plug of the pore at the S5 site. Blocking ion at S5 in conjunction with the hydrophobic repulsion at SF entry creates an ion association or entry barrier to the permeant $K^+$ ion, $2\lambda$ wide in units of fractional electrical distance, Fig 3B, extending through S4 and S3 positions, Fig. A1C.

The different nature of the resulting electrostatics in these two cases implicates different permeation mechanisms and warrants the use of a different biophysical description of unidirectional fluxes. We, therefore, model the unidirectional fluxes as additive over the whole $V_m$ range of the measured Kir4.1 current. In other words, the proposed modeling approach suggests *that in its structure-function a weak rectifier results as a superposition of inward and outward rectifiers*. Single mutation studies [22], [92] suggested their structural determinants are very similar, whereas a single point mutation flips the pore from a strong Kir2.1 to a weak Kir1.1 rectifier.

**Appendix-2**

**Table 2 - Experimental studies reporting altered properties of Kir4.1 current**

| Altered current | Exp. model, Reference | WR-Kir inward | WR-Kir outward | Cell type | Comment |
|---|---|---|---|---|---|
| Kir4.1 current | Astrocytes in ALS [49] | ↓↓ | ↓↓ | Astrocytes in culture | Non-Kir outward current increased markedly in the ALS model |
| Kir4.1 outward current | Astrocytes in Huntington's Disease [48] | ↓ | ↓↓↓ | Astrocytes in slices | |
| Currents through heteromeric Kir4.1/Kr5.1 channels | [45] | Not altered | Abolished | HEK-293 cells | Kir channels introduced by DNA transfection |
| Kir4.1 current | Astrocytes in Depression [93] | ↑↑ | ↑↑↑ | Astrocytes in slices | Not considered in the parametric analysis since the I-V curve remains monotonic. |

**Table 2** – **Selection of studies** where quantitative alteration of the macroscopic Kir current in electro-physiological recordings has been reported, and some form of identification of weakly rectified Kir4.1 or inwardly rectified Kir4.1/Kir5.1 has been elaborated. Many more disease models have reported some *association* of model phenotypes to Kir channels in astrocytes, but differential quantitation of the macroscopic currents or conductances on a single cell has not been among the study objectives.